\documentclass[twocolumn,showpacs,superscriptaddress,preprintnumbers,amsmath,amssymb,prc]{revtex4-2}

\usepackage{graphicx}% Include figure files
\usepackage{dcolumn}% Align table columns on decimal point
\usepackage{bm}% bold math
\usepackage{float}
\usepackage{color}
\usepackage{ulem}
\usepackage{CJK}
\usepackage[colorlinks,linkcolor=blue,urlcolor=blue,citecolor=blue]{hyperref}

\newcommand{\bs}{\boldsymbol}

\usepackage{tikz,xcolor,hyperref}
\definecolor{lime}{HTML}{A6CE39}
\DeclareRobustCommand{\orcidicon}{
\begin{tikzpicture}
\draw[lime, fill=lime] (0,0) 
circle [radius=0.16] 
node[white] {{\fontfamily{qag}\selectfont \tiny ID}};
\draw[white, fill=white] (-0.0625,0.095) 
circle [radius=0.007];
\end{tikzpicture}
\hspace{-2mm}
}
\foreach \x in {A, ..., Z}{%
\expandafter\xdef\csname orcid\x\endcsname{\noexpand\href{https://orcid.org/\csname orcidauthor\x\endcsname}{\noexpand\orcidicon}}
}

\begin{document}
\begin{CJK*}{UTF8}{gbsn}

\title{Lambda polarization in $^{108}$Ag +$^{108}$Ag and $^{197}$Au +$^{197}$Au collisions around a few GeV}
\author{Xian-Gai Deng(邓先概)\orcidA{} }
\affiliation{Key Laboratory of Nuclear Physics and Ion-beam Application (MOE), Institute of Modern Physics, Fudan University, Shanghai 200433, China}
\affiliation{  Shanghai Research Center for Theoretical Nuclear Physics， NSFC and Fudan University, Shanghai 200438, China}

\author{Xu-Guang Huang (黄旭光) \orcidB{}}
\email{Corresponding author: huangxuguang@fudan.edu.cn}
    \affiliation{Department of Physics and Center for Field Theory and Particle Physics, Fudan University, Shanghai, 200433, China}
    \affiliation{Key Laboratory of Nuclear Physics and Ion-beam Application (MOE), Institute of Modern Physics, Fudan University, Shanghai 200433, China}
    \affiliation{  Shanghai Research Center for Theoretical Nuclear Physics， NSFC and Fudan University, Shanghai 200438, China}
\author{Yu-Gang Ma (马余刚)\orcidC{} }
\email{Corresponding author: mayugang@fudan.edu.cn}
    \affiliation{Key Laboratory of Nuclear Physics and Ion-beam Application (MOE), Institute of Modern Physics, Fudan University, Shanghai 200433, China}
    \affiliation{  Shanghai Research Center for Theoretical Nuclear Physics， NSFC and Fudan University, Shanghai 200438, China}

\date{\today}

\begin{abstract}
Within the framework of Ultra-relativistic Quantum Molecular Dynamics (UrQMD) model,  we extract the global spin polarization of $\Lambda$ hyperon in $^{108}$Ag + $^{108}$Ag and $^{197}$Au + $^{197}$Au collisions at $\sqrt{s_{\rm NN}} = 2.42 - 62.4$ GeV. We use two different approaches to calculate the $\Lambda$ polarization $P_y$: approach I is based on  equilibrium assumption so that $P_y$ is determined by thermal vorticity and approach II assumes a proportional relation between $P_y$ and the system's angular momentum in $\Lambda$'s rest frame. We find that both approaches can describe the experimental data at low energies around a few GeV but only approach I describes well also higher-energy data. This suggests that at higher energies the relativistic effect plays an important role. After taking into such effect properly, the relativity-improved approach II can describe the higher-energy data as well.
\end{abstract}


\maketitle

\section{Introduction}
\label{introduction}

Experiments at the Relativistic Heavy Ion Collider (RHIC) have given convincing evidences that the quark-gluon plasma (QGP) is created in relativistic heavy-ion collisions \cite{STAR1_2005,PH_2005}. Also, the RHIC-STAR experiments found that in non-central relativistic heavy-ion collisions the QGP is under high-speed local rotation (or fluid vorticity) with rotating velocity reaching $10^{21}$ s$^{-1}$ \cite{STAR1_2017,STAR:2018gyt}. The particles could thus be spin-polarized through the spin-orbit coupling. In 2005, Liang and Wang proposed that $\Lambda$ (and $\overline{\Lambda}$) hyperons can be polarized along the orbital angular momentum of two colliding nuclei \cite{LZT_2005}. For the Beam Energy Scan (BES) program at RHIC, it is found that the global spin polarization of $\Lambda$ and $\overline{\Lambda}$ hyperons are different, 
especially at $\sqrt{s_{\rm NN}}$ = 7.7 GeV \cite{STAR1_2017,STAR:2018gyt}. One of reasons for this difference is probably the magnetic field effect \cite{Becattini:2016gvu,GY_2019}, but there are other possible causes for this difference \cite{SYF17,HZZ_2018,OV_20,Csernai:2018yok,Xie:2019wxz}. Another feature is that both  $\Lambda$ and $\overline{\Lambda}$ global polarization decrease with increasing collision energy $\sqrt{s_{\rm NN}}\geq 7.7$ GeV which is consistent with the collision energy dependence of fluid vorticity at mid-rapidity \cite{Deng:2016gyh,Jiang:2016woz}. The calculations from hydrodynamical \cite{Shen:2020mgh,Wu:2021xgu} and transport-model approaches can explain well the experimental results for global $\Lambda$ polarization  \cite{KL_2017,LH17,XYL_2020,YBI_2020,Fu:2020oxj,Shi:2017wpk,Xia:2018tes,Wu:2019eyi,WDX19} and also $\Xi$ and $\Omega$ polarization \cite{WDX19,STAR:2020xbm,Li:2021zwq}, eg., see Refs.~\cite{Huang:2020xyr,FB_2020,Karpenko:2021wdm,Huang:2020dtn,Gao:2020vbh,Liu:2020ymh} for reviews. 

Hyperon spin polarization below $\sqrt{s_{\rm NN}}$ = 7.7 GeV is less explored. Some experiments, such as the STAR fixed target program at RHIC ($\sqrt{s_{\rm NN}}$ = 3$-$7.2 GeV) \cite{STAR2_2017}, the HADES at GSI Helmholtzzentrum f{\"u}r Schwerionenforschung ($\sqrt{s_{\rm NN}}$ = 2.3$-$2.6 GeV) \cite{HADES_2009}, the Nuclotron based Ion Collider fAcility (NICA) in Dubna ( $\sqrt{s_{\rm NN}}$ = 4$-$11 GeV) \cite{VD17}, the Baryonic Matter at Nuclotron (BM@N) in Dubna ($\sqrt{s_{\rm NN}}$ = 2.3$-$3.5) \cite{MK19}, Facility for Antiproton and Ion Research (FAIR) in Darmstadt ($\sqrt{s_{\rm NN}}$ = 2.7$-$4.9 GeV) \cite{TA17}, the High Intensity heavy-ion Accelerator Facility (HIAF) in Huizhou, China ($\sqrt{s_{\rm NN}}$ = 2.3$-$4 GeV) \cite{LL17,Liu,Niu}, can reach such energy region. In Ref. \cite{DXG20}, it was predicted that both the kinematic vorticity and thermal vorticity at mid-rapidity show non-monotonic behavior as functions of $\sqrt{s_{NN}}$ and their maximum values are reached about $\sqrt{s_{\rm NN}}\sim$ 3 GeV. This may suggest a possible similar behavior for $\Lambda$ polarization. Very recently, the STAR and HADES Collaborations reported their measurements of global $\Lambda$ polarization for $^{197}$Au +$^{197}$Au collisions at $\sqrt{s_{\rm NN}}$ = 3 and 2.4 GeV \cite{STAR:2021beb,HADES:2021}, respectively, and for $^{108}$Ag + $^{108}$Ag collisions at $\sqrt{s_{\rm NN}}$ = 2.55 GeV \cite{HADES:2021}. The purpose of the this paper is to extend the previous calculations for kinematic and thermal vorticities in Ref. \cite{DXG20} to extract the spin polarization of $\Lambda$ hyperons at energies $\sqrt{s_{\rm NN}}$ = 2.42 $-$ 62.4 GeV with two different methods to connect the spin polarization to the vorticities or angular momentum of the medium. We will also compare our results with the recent calculations in Refs. \cite{YBI21,Guo:2021uqc,Ayala:2021xrn,Ivanov:2019ern}.

%In 2019, the measurement of $\Lambda$  polarization at $\sqrt{s_{\rm NN}}$ = 2.4 GeV was reported by the HADES Collaboration and the value is close to zero \cite{HADES:2019}. However, for recent result from HADES \cite{HADES:2021}, polarization at $\sqrt{s_{\rm NN}}$ = 2.4 GeV for $^{197}$Au+$^{197}$Au collisions is given about 5\%. Thus the behavior of global polarizations at the energy region of $\sqrt{s_{\rm NN}}$ = 2.4 $-$ 7.7 GeV is still not clear. As prediction, in Ref. \cite{DXG20}, ones found the vorticity and thermal vorticity at mid-rapidity region show maximum values around  $\sqrt{s_{\rm NN}}$ = 3$-$5GeV and the energy position depends on the centrality~\cite{DXG20} which is insistent with the result within the model of three-fluid dynamics (3FD) \cite{YBI21}. The purpose of this paper is to extract polarization of $\Lambda$ hyperons with two methods at free-out point of $\Lambda$ and research more acknowledge about polarization behavior around $\sqrt{s_{\rm NN}}$ = 3.0 GeV. 

\section{Setup of the calculation}
\label{entropyandmodel}

The global spin polarization may reflect the medium's ability of transferring the global orbital angular momentum to spin angular momentum. Thus, the strength of global spin polarization depends substantially on the dynamics of spin-orbit coupling and the medium properties (which determines how much and in what form the orbital angular momentum is retained in the medium).  Therefore, in general it is a difficult task to model the global spin polarization of hyperons. In the following, we will use two different approaches to model the global $\Lambda$ polarization. 

{\bf $\bullet$ Approach I.} The first approach assumes global equilibrium of spin degree of freedom so that the spin polarization is fully determined by thermal vorticity. In this case, the mean spin vector of spin-1/2 fermion with on-shell momentum $p^\mu$ is given by \cite{Becattini:2013fla,Fang:2016vpj,Fang2,Liu:2020flb},
\begin{equation}
S^{\mu} (x,p) = -\frac{1}{8m}(1-n_F)\epsilon^{\mu\nu\rho\sigma}p_{\nu}\varpi_{\rho\sigma}(x),             
\label{SPINTENSOR-0}
\end{equation}
where $n_F = n_F(x,p)$ is the Fermi-Dirac function and $\varpi_{\rho\sigma}$ is the thermal vorticity tensor:
\begin{equation}
\varpi_{\mu\nu}=\frac{1}{2}(\partial_{\nu}\beta_{\mu}-\partial_{\mu}\beta_{\nu}).                    
\label{thvorticity}
\end{equation}
In Eq.~(\ref{thvorticity}), $\beta^\mu = \beta u^\mu$ with $\beta = 1/T$ the inverse temperature and $u^\mu = \gamma(1,\bs{v})$ is the fluid four-velocity with $\gamma = 1/\sqrt{1-\boldsymbol{v}^2}$ the Lorentz factor. One can decompose the thermal vorticity in Eq.~(\ref{thvorticity}) as,
\begin{eqnarray}
\bs{\varpi}_{T}&=&\frac{1}{2}\Big{[} \nabla \Big{(}\frac{\gamma}{T}\Big{)}+\partial_{t} \Big{(}\frac{\gamma \bs{v}}{T}\Big{)}\Big{]},
\label{thvorticity-1} \\
\bs{\varpi}_{S}&=&\frac{1}{2} \nabla \times \Big{(}\frac{\gamma \bs{v}}{T} \Big{)},
\label{thvorticity-2}
\end{eqnarray}
where $\bs{\varpi}_{T} $ and $\bs{\varpi}_{S}$ are called `$T$' and `$S$' thermal vorticity, respectively. For $\Lambda$ hyperon, $n_F \approx 0$, because its yield in each collision is small, and thus Eq.~(\ref{SPINTENSOR-0}) with Eqs.~(\ref{thvorticity-1}) and (\ref{thvorticity-2}) gives,
\begin{eqnarray}
S^{0}({x},{\bf p})&=&\frac{1}{4m}{\bf p}\cdot \bs{\varpi}_{S},\\
\label{SPINTENSOR-1}
\bs{S}({x},{\bf p})&=&\frac{1}{4m}  \Big{(} E_{p}\bs{\varpi}_{S}+{\bf p}\times \bs{\varpi}_{T}   \Big{)},
\label{SPINTENSOR-2}
\end{eqnarray}
where $E_{p}, {\bf p}$ and $m$ are the $\Lambda$'s energy, momentum, and mass in the center of mass (c.m.) frame of nucleus-nucleus collision. Since the polarization of $\Lambda$ in experiment is measured in the rest frame of $\Lambda$, one needs to transform Eq.~(\ref{SPINTENSOR-2}) from c.m. frame of nucleus-nucleus collision to the rest frame of $\Lambda$ by a Lorentz boost,
\begin{equation}
\bs{S}^{\ast}({x},{\bf p}) = \bs{S}-\frac{{\bf p} \cdot \bs{S}}{E_{p}(m+E_{p})}{\bf p}. 
\label{SPINTENSOR-3}
\end{equation}
Once the thermal vorticity at the freeze-out point of each $\Lambda$ is known, we can obtain the corresponding $\bs{S}^{\ast}(x,{\bf p})$. The averaged spin vector is given by (using energy density $\epsilon({x})$ as a weight which is given below Eq.~(\ref{edensity})), 
\begin{equation}
\langle \bs{S}^{\ast} \rangle=\frac{\sum_{a}^{N} \bs{S}^{\ast}({x}_{a},{\bf p}) \epsilon({x}_{a})}{\sum_{a}^{N} \epsilon({x}_{a}) },            
\label{SPINTENSOR-31}
\end{equation}
where the summation is over all $\Lambda$'s within a given kinematic region and $x_a$ is the freeze-out coordinate of the $a$th $\Lambda$. Then the polarization of $\Lambda$ in the three-direction $\bs{n}$ is given by
\begin{equation}
P_{\bs{n}} = 2 \langle \bs{S}^{\ast} \rangle \cdot \bs{n}.                    
\label{SPINTENSOR-4}
\end{equation}

{\bf $\bullet$ Approach II.} The second approach does not assume global equilibrium but assumes a simple proportional relation between the spin vector and the orbital angular momentum of the matter, reflecting the fact that the global $\Lambda$ polarization is due to the global angular momentum no matter whether the system is at equilibrium or not. Thus, for a $\Lambda$ hyperon of momentum $\bf p$ frozen-out at coordinate $x$, we assume its spin vector in its rest frame to be given by
\begin{equation}
\bs{S}^\ast({x},{\bf p})=\frac{\chi}{2} \bs{J}^\ast(x),
\label{SPINapproach2}
\end{equation}
where $\bs{J}^\ast(x)$ is the orbital angular momentum relative to $\Lambda$ at freeze-out coordinate $x$ and $\chi$ is a susceptibility that characterized the matter's ability of transferring orbital angular momentum to spin. In principle, the value of $\chi$ should depend on the equation of state and spin-orbital coupling of the matter and thus may change with collision energy or centrality. Due to the lack of the knowledge of $\chi$, in our simulation, we simply assume a constant value, $\chi$ = 1/20, as it can describe the experimental data the best. The relative orbital angular momentum $\bs{J}^\ast(x)$ is calculated through
\begin{equation}
\bs{J}^\ast(x)=\frac{\sum_i ({\bf x}_i-{\bf x})\times{\bf p}^\ast_i\rho(x,x_i)}{\sum_i\rho(x,x_i)},
\label{angmom}
\end{equation}
where the summation is over the participants, ${\bf p}^\ast_i={\bf p}_i-{\bf p}$ is the relative momentum of the $i$th particle with respect to the rest frame of $\Lambda$ frozen-out at $x$, and $\rho(x,x_i)$ is the smearing function. Since $\rho(x,x_i)$ is very localized (we will choose it as a Gaussian, see Eq.~(\ref{edensity})), the summation is essentially over particles in a small volume around $\bf{x}$; and thus the physical meaning of $\bs{J}^\ast(x)$ is the orbital angular momentum per particle in a small volume around $\bf{x}$ as specified by $\rho(x,x_i)$. Once $\bs{S}^\ast(x,{\bf p})$ is obtained, the $\Lambda$ polarization for given $\bf{p}$ is calculated also using Eq.~(\ref{SPINTENSOR-4}) which leads to $P_{\bs n}=\chi\langle\bs{J}^\ast\rangle\cdot\bs{n}$ where the average is given in same manner of Eq.~(\ref{SPINTENSOR-31}). Note that the approach II is phenomenological so that the susceptibility $\chi$ is fitting parameter. Besides, unlike approach I, the approach II is non-relativistic. This is expected to be not a problem at low energies, but it may not work at high energies where the relativistic effects are important; we will come to this point in the next section.

In this work, following closely Ref. \cite{DXG20}, simulations are performed in the framework of Ultra-relativistic Quantum Molecular Dynamics (UrQMD) model which includes particle re-scattering, color string fragmentation, and formation and decay of hadronic resonances~\cite{SA98,MB99,HP08,JS18}. A slight difference from Ref. \cite{DXG20} is that we now use a smearing function in which a Lorentz contraction factor $\gamma_z$ along the beam direction is taken into account \cite{HP08}:
\begin{equation}
\begin{split}
 &\rho(x,x_i)=  \\
 &\frac{\gamma_{z}}{(2\pi\sigma^2)^{3/2}}\,{\rm exp} \Big{\{} -\frac{({\rm x}-{\rm x}_{i})^{2}+({\rm y}-{\rm y}_{i})^{2}+[\gamma_{z}({\rm z}-{\rm z}_{i})]^{2}}{2\sigma^{2}}\Big{\}},                  
\end{split}
\label{edensity}
\end{equation}
%In the UrQMD, each particle is described by a three-dimensional Gaussian distribution of its total energy, momentum in $x, y$, and $z$ directions. By taking into account Lorentz contraction of the nuclei in the longitudinal  direction, a Lorentz factor $\gamma$ is used \cite{HP08}. Then the energy density in the c.m. frame of nucleus-nucleus collisions can be given by where $A= (1/2\pi)^{3/2}\gamma_{z}/\sigma^{3}E_{c.m.nn}$ is  normalization factor, $\epsilon_{c.m.nn}$ and $E_{c.m.nn}$ are the energy density and total energy of particle in c.m. frame of nucleus-nucleus collisions, respectively. And for the width of Gaussian $\sigma$, we choose to be $\sigma = 1.48$ fm for baryons~\cite{Hartnack:1997ez} and $\sigma = 0.98$ fm for mesons from a constituent quark number scaling for volume $\sigma_{ \pi} = (2/3)^{1/3}\sigma_{\rm p,n}$ \cite{DXG20}. The ``temperature" $T({\bf x})$ simply as a measure of the energy density, $\epsilon ({\bf x})$, via the relation $\epsilon = a T^{4}$; The prefactor $a$ is chosen here to be $a = \pi^{2}(16+10.5N_{f})/30\approx 15.6$ with $N_{f} = 3$. 
where the width of Gaussian $\sigma$ is chosen to be $\sigma = 1.48$ fm for baryons~\cite{Hartnack:1997ez} and $\sigma = 0.98$ fm for mesons from a constituent quark number scaling for volume $\sigma_{ \pi} = (2/3)^{1/3}\sigma_{\rm p,n}$ \cite{DXG20}. The temperature $T({x})$ is obtained from the energy density and the number density, 
\begin{equation}
\epsilon ({ x})=\sum_{i}E_{i}\rho(x,x_i), \,\,\, \rho ({ x})=\sum_{i}\rho(x,x_i), \nonumber 
\end{equation}
where $E_{i}$ the energy of the $i$th particle and the summation over all the nucleons and pions. Re-expressing the energy density ($\epsilon$) and the number density ($\rho$) by the nucleon and pion distribution functions,
\begin{eqnarray}
\epsilon &=&\int f_{n}E_{n}d^{3}p+\int f_{pion}E_{pion}d^{3}p,\\
\label{fdis-1}
\rho &=& \int f_{n}d^{3}p+\int f_{pion}d^{3}p,
\label{fdis-2}
\end{eqnarray}
where $f_{n}=\frac{4}{(2\pi\hbar)^{3}}\frac{1}{\exp[\frac{E_{n}-\mu_{n}}{T}+1]}$ is the distribution function of nucleons and $f_{pion}=\frac{3}{(2\pi\hbar)^{3}}\frac{1}{\exp[\frac{E_{pion}}{T}-1]}$ is the distribution function of pions, one can extract the nucleon chemical potential $\mu_n$ and the temperature $T$. Note that the chemical potential for pions is treated as zero. We note that the matter after the collisions at low energies  may not reach the local equilibrium \cite{ZhangGQ} so that the temperature and the chemical potential here are not qualified as thermodynamic quantities but merely a parametrization of the energy density and the number density.

The velocity field is calculated in the same way as described in Ref.~\cite{DXG20}:
\begin{equation}
\bs{v}(x)=\frac{\sum_{i} ({\bf p}_i/E_i)\rho(x,x_i)}{\sum_{i} \rho(x,x_i) }.\end{equation}

\section{Results and discussion}
\label{resultsSH}
\begin{figure}[htb]
%\begin{figure}
\setlength{\abovecaptionskip}{0pt}
\setlength{\belowcaptionskip}{8pt}
\includegraphics[scale=1.10]{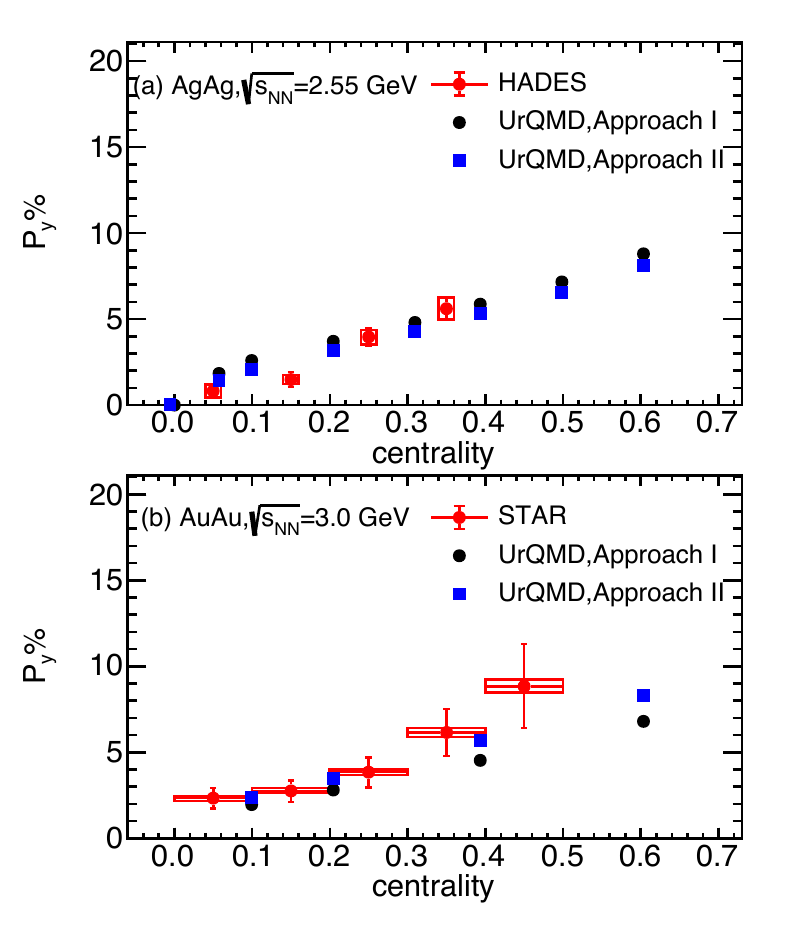}
\caption{(Color online) Global $\Lambda$ polarization as a function of centrality  for (upper panel) $^{108}$Ag+$^{108}$Ag collisions for 0.2$<p_{T}$ [GeV/c]$<$1.5 and $|{\rm Y}|<$0.3 at $\sqrt{s_{\rm NN}}=2.55$ GeV and (lower panel) $^{197}$Au+$^{197}$Au collision for $p_{T}$ $>$0.7 GeV/c and $-0.2<{\rm Y}<1$ at $\sqrt{s_{\rm NN}}$ = 3 GeV. The HADES data for $\Lambda$ polarization at 0.2$<p_{T}$ [GeV/c]$<$1.5 and $-0.5<{\rm Y}<0.3$ \cite{{HADES:2021}} and the STAR data at $p_{T}$ $>$0.7 GeV/c and $-0.2<{\rm Y}<1 $ \cite{STAR:2021beb} are also shown.}
\label{fig:fig1}
\end{figure}

The global $\Lambda$ polarization as a function of centrality is given in Fig.~\ref{fig:fig1} based respectively on the two different approaches described in the last section. Here global polarization  $P_y$ is calculated by averaging over all $\Lambda$'s with rapidity $|{\rm Y}|<0.3$ and $0.2$ GeV$< p_T<1.5$ GeV in $^{108}$Ag+$^{108}$Ag collisions and with rapidity $-0.2<{\rm Y} < 1$ and $p_T>0.7$ GeV in  $^{197}$Au+$^{197}$Au collisions. The centrality is determined through $C = b^{2}/b^{2}_{max}$, with $b_{max} = R_{T}+R_{P} = 1.2A_{T}^{1/3}+1.2A_{P}^{1/3}$ with $A_{T/P}$ the atomic number of target/projectile nucleus. The polarization calculated based on both approaches I and II fit to the experimental datas from HADES Collaboration \cite{HADES:2021} and STAR Collaboration \cite{STAR:2021beb}. The global $\Lambda$ polarization increases with centrality reflecting the fact that the system's angular momentum increases with centrality. Note that at high energies the global polarization show similar centrality dependence which can also be described by approach I (see, e.g., Ref. \cite{Fu:2020oxj}). 

\begin{figure}[htb]
%\begin{figure}
\setlength{\abovecaptionskip}{0pt}
\setlength{\belowcaptionskip}{8pt}
\includegraphics[scale=0.5]{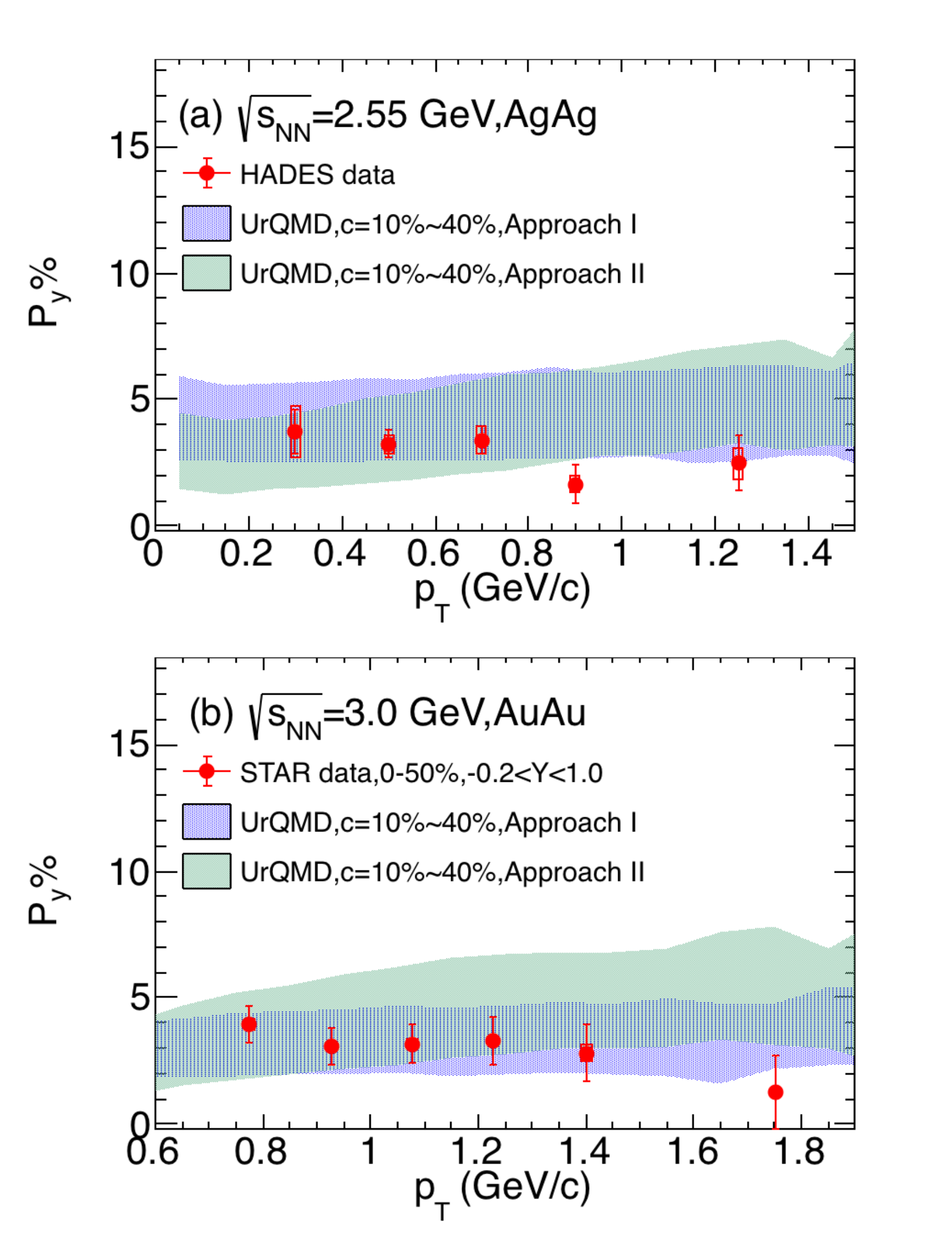}
\caption{(Color online) $\Lambda$ polarization as a function of transverse momentum $p_{T}$ for (upper panel) $^{108}$Ag+$^{108}$Ag collision for $|{\rm Y}|<0.3$ at $\sqrt{s_{\rm NN}}$ = 2.55 GeV and (lower panel) $^{197}$Au+$^{197}$Au collision for $-0.2<{\rm Y}<1$ at $\sqrt{s_{\rm NN}}$ = 3 GeV. The red solid dots are  experimental results for $^{108}$Ag+$^{108}$Ag collision for -0.5$<$Y$<0.3$  \cite{HADES:2021} and $^{197}$Au+$^{197}$Au collision for $-0.2<{\rm Y}<1$ \cite{STAR:2021beb}.}
\label{fig:fig2}
\end{figure}
\begin{figure}[htb]
%\begin{figure}
\setlength{\abovecaptionskip}{0pt}
\setlength{\belowcaptionskip}{8pt}
\includegraphics[scale=0.5]{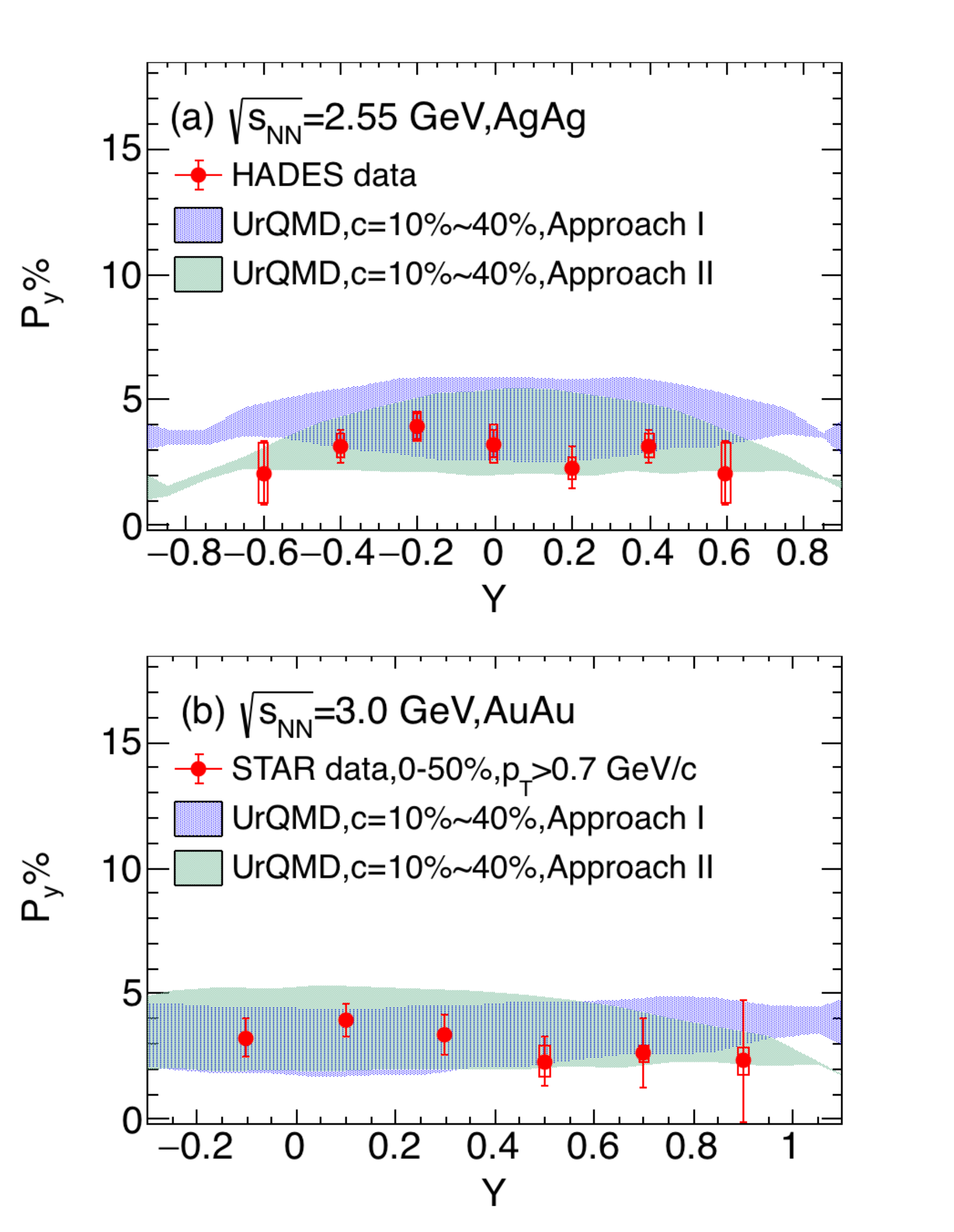}
\caption{(Color online) $\Lambda$ polarization as function of rapidity ${\rm Y}$ for $^{108}$Ag+$^{108}$Ag collision for 0.2$<p_{T}$ [GeV/c]$<$1.5 at $\sqrt{s_{\rm NN}}$ = 2.55 GeV and $^{197}$Au+$^{197}$Au collision for $p_T>0.7$ GeV/c at $\sqrt{s_{\rm NN}}$ = 3 GeV. The red solid dots are experimental results for 0.2$<p_{T}$ [GeV/c]$<$1.5 in $^{108}$Ag+$^{108}$Ag collision \cite{HADES:2021} and $p_T>0.7$ GeV/c in $^{197}$Au+$^{197}$Au collision.}
\label{fig:fig3}
\end{figure}

Further, in Fig.~\ref{fig:fig2}, we show $\Lambda$ polarization as a function of transverse momentum $p_{T}$ for $^{108}$Ag+$^{108}$Ag collision at $\sqrt{s_{\rm NN}}$ = 2.55 GeV within $10\%\sim40 \%$ centrality and for $^{197}$Au+$^{197}$Au collision at $\sqrt{s_{\rm NN}}$ = 3 GeV within $10\%\sim40 \%$ centrality and compare to the recent experimental results reported by HADES Collaboration \cite{HADES:2021} and by STAR Collaboration \cite{STAR:2021beb}. The results of both approaches I and II are shown with blue and green bands, respectively. It is found that both approaches can give reasonable description of the data, although the approach II gives a $\Lambda$ polarization slightly increasing with $p_T$ and overestimates the data at high $p_T$. Similar increasing pattern is also found in Ref. \cite{Guo:2021uqc} around the same energies and at higher energies using parton cascade model (the AMPT model) based on approach I (see, e.g., Ref.~\cite{WDX19}), but different from the results in Ref.~\cite{SYF17} based on chiral kinetic theory and in Ref.~\cite{Fu:2020oxj} based on a hydrodynamical model at higher energies, making the theoretical results quite model dependent in describing the $p_T$ dependence. The dependence of $\Lambda$ polarization on rapidity is shown in  Fig.~\ref{fig:fig3}. One can see that both of the two approaches are around the experimental results at the rapidity region of -0.6 $<$Y$<$ 0.6 for $^{108}$Ag+$^{108}$Ag collision and -0.2 $<{\rm Y}<$ 1 for $^{197}$Au+$^{197}$Au collision. %Polarization decreases around $|Y| \leq 0.8$.  Combined from Fig.~\ref{fig:fig1} to Fig.~\ref{fig:fig4}, it implies that the global polarization of $\Lambda$ hyperons tends to the `$T$'  thermal vorticity as in Eq.(\ref{thvorticity-1}).

\begin{figure}[htb]
%\begin{figure}
\setlength{\abovecaptionskip}{0pt}
\setlength{\belowcaptionskip}{8pt}
\includegraphics[scale=0.38]{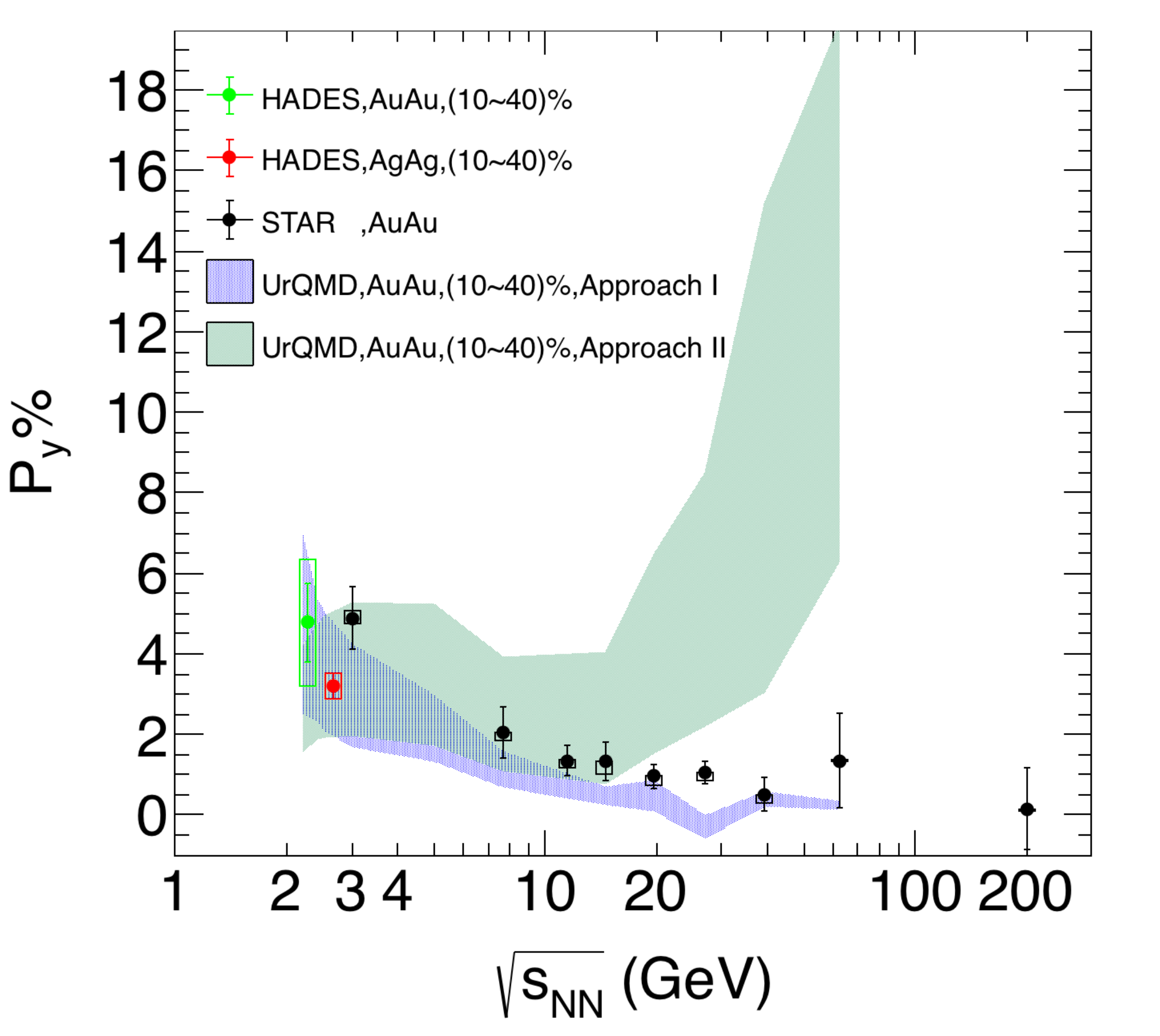}
\caption{(Color online) Global $\Lambda$ polarization as a function of collision energy $\sqrt{s_{\rm NN}}$. The red and green solid dots are from the HADES \cite{HADES:2021} and STAR data for $\Lambda$ global polarization are displayed with black solid dots \cite{STAR1_2017,STAR:2021beb}. The UrQMD calculations is for $^{197}$Au+$^{197}$Au collision at 0.4 $<p_{T}$ [GeV/c]$<$  3.0 and $|{\rm Y}|<$0.3.}
\label{fig:fig4}
\end{figure}

\begin{figure}[htb]
%\begin{figure}
\setlength{\abovecaptionskip}{0pt}
\setlength{\belowcaptionskip}{8pt}
\includegraphics[scale=0.38]{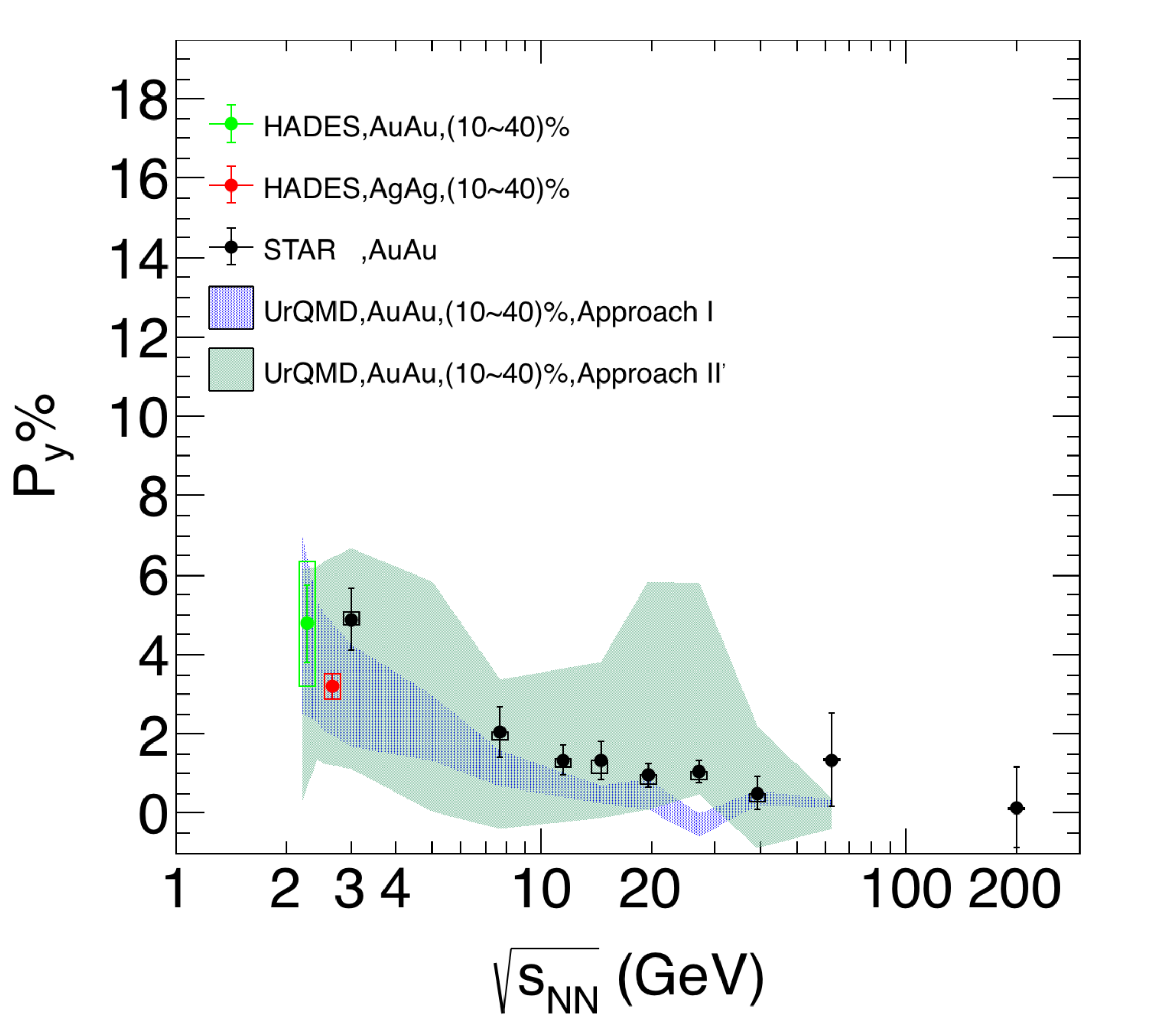}
\caption{(Color online) The same as  Fig.~\ref{fig:fig4} but with the approach II replaced by a relativity-improved one (the approach II').}
\label{fig:fig5}
\end{figure}

Finally,  the collision energy dependence of global $\Lambda$ polarization is shown in Fig.~\ref{fig:fig4}. In our UrQMD calculation, we focus on the global polarization of $\Lambda$ in $^{197}$Au+$^{197}$Au collision at the energies of 2.5 GeV $<\sqrt{s_{\rm NN}}<$ 62.4 GeV. The $\Lambda$ polarization in $^{108}$Ag+$^{108}$Ag collision is expected to be larger than that in $^{197}$Au+$^{197}$Au at the same energy because the former system has smaller size \cite{Shi:2017wpk}. The $\Lambda$ global polarization via approach I is shown by the blue band and via approach II is displayed with green band in which a maximum value around  $\sqrt{s_{\rm NN}}$ = 4 GeV is seen. 

Both approaches can fit to the experimental data around $2.5-10$ GeV within the error bars. As at $\sqrt{s_{\rm NN}}=2m_{\rm N}$ we do not expect any finite global polarization, we would thus expect the existence of a maximal global polarization around a few GeV. If such a maximum exists, approach I suggests that it appears below 2.5 GeV while approach II suggests the maximum to be around 3-5 GeV. This is consistent with the initial kinematic and thermal vorticities which have also a maximum value around 3$-$5 GeV~\cite{DXG20} depending on the centrality. We note that the appearance of a maximum in the global $\Lambda$ polarization versus $\sqrt{s_{\rm NN}}$ is well expected from the appearance of the maximum in vorticities and was also discussed in Refs.~\cite{YBI21,Guo:2021uqc,Ayala:2021xrn}; but the prediction for the location of the maximum is model dependent. It will be of great interest if the experiments can test whether such a maximum exists and, if it exists,  where it is located.

At energies around $10 - 62.4$ GeV, approach I still describes well the experimental data of global $\Lambda$ polarization. For the approach II, however, the numerical result begins to increase as energy $\sqrt{s_{\rm NN}}$ becomes larger than $\sim 15$ GeV. The reveal the reason for such a behavior, we notice that the formula (\ref{angmom}) for approach II is a non-relativistic one, which means that it should be valid only at low energies where the relativistic effect is not important. Due to the lack of relativistic correction, it diverges at higher energies (when the momenta of the particles ${\bf p}_i^*$ become larger and larger). To recover a relativistically corrected formula for approach II (labeled as approach II'), we use the following procedure. First, calculate the angular momentum ${\bs J}(x)$ in the laboratory frame (i.e., the c.m. frame of the collision) through 
\begin{equation}
\bs{J}(x)=\frac{\sum_i ({\bf x}_i-{\bf x})\times{\bf p}_i\rho(x,x_i)}{\sum_i\rho(x,x_i)},
\label{angmom-1}
\end{equation}
where $x$ is coordinate at which the $\Lambda$ (with momentum ${\bf p}$) is frozen out. Then, the spin vector in the laboratory frame is given by,
\begin{equation}
\bs{S}({x})=\frac{\chi}{2} \bs{J}(x).
\label{SPINapproach2-1}
\end{equation}
Finally, the spin polarization in the rest frame of $\Lambda$ is given through Eqs.~(\ref{SPINTENSOR-3}) - (\ref{SPINTENSOR-4}). The numerical result of approach II' is given in Fig.~{\ref{fig:fig5}}. Overall, it can describe the trend of the experiment data well in all the energies we have considered. But one can notice a bump around $\sqrt{s_{\rm NN}}$=20 GeV which is due to the numerical fluctuation in the UrQMD model itself.

\section{Conclusions}
\label{summary}

In summary, within the framework of the UrQMD model, we extract the global polarization of $\Lambda$ hyperon with two approaches. One with thermal equilibrium assumption and another is via the relative angular momentum in $\Lambda$'s rest frame. Our simulation shows that both approaches can describe the experimental data around $\sim 2.42-10$ GeV. At these energies, we do not expect that the spin degree of freedom reach equilibrium, but surprisingly even the approach I which is based on equilibrium assumption could explain the experimental results well. At higher energies, approach II deviates from the experimental data severely due to the lack of relativistic effect. We thus check that once the relativistic correction is properly included through the Lorentz transformation, the modified approach II can also describe higher-energy data. Our result suggests that the global polarization is a result of the global angular momentum of the system so that it is insensitive to whether the intermediate stages are at equilibrium or not. Thus it would be very interesting to test the approaches I and II (including the relativity-improved one) by studying the local spin polarization, namely, the transverse and longitudinal $\Lambda$ polarization at different azimuthal angles~\cite{Huang:2020xyr,FB_2020,Karpenko:2021wdm,Huang:2020dtn,Gao:2020vbh,Liu:2020ymh}. Besides, in this work, we do not consider the feed-down decay from more massive hyperons which could reduce the primary $\Lambda$ polarization by about $10-15 \%$ \cite{Xia:2019fjf,Becattini:2019ntv}. The phenomenological parameter $\chi$ in approach II may be studied theoretically at low energies through kinetic methods~\cite{Peng:2021ago,Gao:2022gqr}. We will leave such theoretical studies to future work. We will also study the global $\Lambda$ polarization in isobar collisions~\cite{Wang:2018ygc,STAR:2019bjg,STAR:2021mii} which may help to understand whether the magnetic field is the source for the splitting of $\Lambda$ and $\overline{\Lambda}$ polarization.

\begin{acknowledgments}
We thank J. Adams, T. Galatyuk, M. Lisa, T. Niida for helpful communications. This work has received partial support from the Guangdong Major Project of Basic and Applied Basic Research No. $2020$B$0301030008$,  the National Natural Science Foundation of China under Contract No.\ $11890714$, No. $11891070$, No. $12075061$, and No. $12147101$, the Key Research Program of the CAS under Grant No. XDB34000000, China Postdoctoral Science Foundation Grant No.\ 2019M661332 and  Postdoctoral Innovative Talent Program of China No.\ BX20200098, and Shanghai Natural Science Foundation through Grant No.~20ZR1404100.
\end{acknowledgments}

\end{CJK*}
\end{document}